\documentclass{PoS}

\title{Super KEKB and Belle II: Status of the KEK Super B Factory}

\ShortTitle{Super KEKB and Belle II}

\author{\speaker{Zden\v{e}k Dole\v{z}al}\thanks{For the Belle II collaboration}\\
        Charles University in Prague, Faculty of Mathematics and
        Physics, \\ 
V Holesovickach 2, Prague, The Czech Republic CZ180 00\\
        E-mail: \email{Zdenek.Dolezal@mff.cuni.cz}}

\abstract{
The Belle detector at the KEKB electron-positron collider has collected  
approximately 800 million $\Upsilon(4\mathrm{S})$ events in its decade of operation.  
The KEKB group has proposed Super-KEKB, an upgrade of KEKB to increase  
the luminosity by two orders of magnitude during a three-year shutdown,  
with an ultimate goal of $8\times 10^{35}\mathrm{cm}^{-2}\mathrm{s}^{-1}$ luminosity. 

To exploit the  
increased luminosity, an upgrade of the Belle detector has been proposed.  
A new international collaboration Belle-II, is being formed, with a  
broader participation of European institutes. Super-KEKB and Belle-II  
were officially placed on the KEK 5-year Roadmap in early 2008. The paper  
presents physics motivation, basic methods of the accelerator upgrade, as well as key  
improvements of the detector. More details are given on the DEPFET  
pixel detector that will be completely built in Europe. 
}

\FullConference{European Physical Society Europhysics Conference on High Energy Physics,
EPS-HEP 2009,\\
		 July 16 - 22 2009\\
		 Krakow, Poland}

\begin{document}

\section{Introduction}
Studying phenomena with heavy flavour involved played a crucial role
in forming Standard Model (SM). Past decade brought exciting discoveries
and new measurements namely at the processes with B-mesons produced.
These experiments were carried out at
two B-factories (PEP III at SLAC and KEKB at KEK) where asymmetric
beams of e$^+$e$^-$ collide at the    $\Upsilon$ resonance which
predominantly decays to B$\bar{\mathrm{B}}$.

The detectors built at the B-factories registered an enormous number of
1.5 billion  B$\bar{\mathrm{B}}$ pairs. BaBar at PEP III (stopped in 2008) collected 553
fb$^{-1}$ integrated luminosity, while  Belle at KEK is still in
operation and has accumulated 950 fb$^{-1}$ until July 2009.

A few examples of the scientific highlights of these factories are
\vspace{-2mm}
\begin{itemize}
\item discovery of CP violation in B system
\vspace{-3mm}
\item measurement of CKM matrix elements
\vspace{-3mm}
\item observation of new charmonium and charmonium-like states 
\vspace{-3mm}
\item discovery of D$^0$ mixing
\vspace{-3mm}
\item many probes of New Physics
\end{itemize}
\vspace{-2mm}

\section{Physics Potential of the Super B factory}
Despite the success of the current B-factories many
questions yet remain to be answered. Many of the existing measurements
have statistical uncertainties higher than their systematics.
More statistics would bring a substantial improvement in the accuracy.
The most prominent are the parameters of CKM matrix, especially its
complex phase, related to the angles of the    CKM unitarity triangle.
%namely the phases (angles of CKM triangle). 
Better determination of
$\phi_2=\alpha$ would be possible: it is strongly constrained
by the $B^0\rightarrow \rho^0 \rho^0$,  $B^0\rightarrow \pi^0
\pi^0$ and $B^0\rightarrow \rho^0 \pi^0$ decay channels. With 5
ab$^{-1}$ $\phi_2$ can be measured with an uncertainty of 2 degrees.
The value of $\sin 2 \phi_1= \sin 2\beta$ can be achieved with a precision of 1\% with 5 ab$^{-1}$.
         Here theory uncertainty is also ~1\%. Third angle
         $\phi_3=\gamma$ can be measured at least with 5 degree error with         5 ab$^{-1}$.

Precise determination of CKM matrix would yield an important input for
SM as well as for New Physics theories. With this precision any
discrepancy with the KM scheme will provide insight into NP. 

This is not the only way NP can be observed. Many of the
studied channels are penguin processes, highly suppressed  in the
Standard Model. Also CP violation in these processes should be small.
Analysis of current Belle and BaBar data show several hints of
discrepancies between SM and measured quantities. With statistics
increased by two orders of magnitude statistically significant effects
could be established. 

Let us recall a well known fact from the kaon era: discrepancies
between theory and experiment can be a very powerful probe to much
higher energy scale. Within the Minimal Flavour Violation scenario, for
example, the current limits on the NP contribution to FCNC translate
into the mass scale of NP above 100 GeV. Super B factory with the
greatly enhanced sensitivity will push this limit by an order of
magnitude higher.

These were only a few ideas about physics potential of the upgraded KEKB.
More details can be found at~\cite{phys5,phys6}. 
 \section{Accelerator Upgrade}
\label{sec:acc}
The project of Super-B factory in in Japan is based on the upgrade of
existing KEKB collider. It is an asymmetric machine using beam of
electrons of 8~GeV and positrons of 3.5~GeV. The collider operates
since 1999 at the High Energy Accelerator Research Organization KEK in Tsukuba. Its
parameters have been constantly improved throughout past decade and
now the machine operates in an excellent mode, delivering the highest
luminosity in the world $\mathcal{L}_{\mathrm{max}} = 2.1 \times 10^{34}\mathrm{cm}^{-2}\mathrm{s}^{-1}$. This has been possible by a crab
cavity, installed in 2007, and by other improvements. 

\begin{figure}
  \centering
\includegraphics[width=12cm,bb=70 60 700 406,clip]{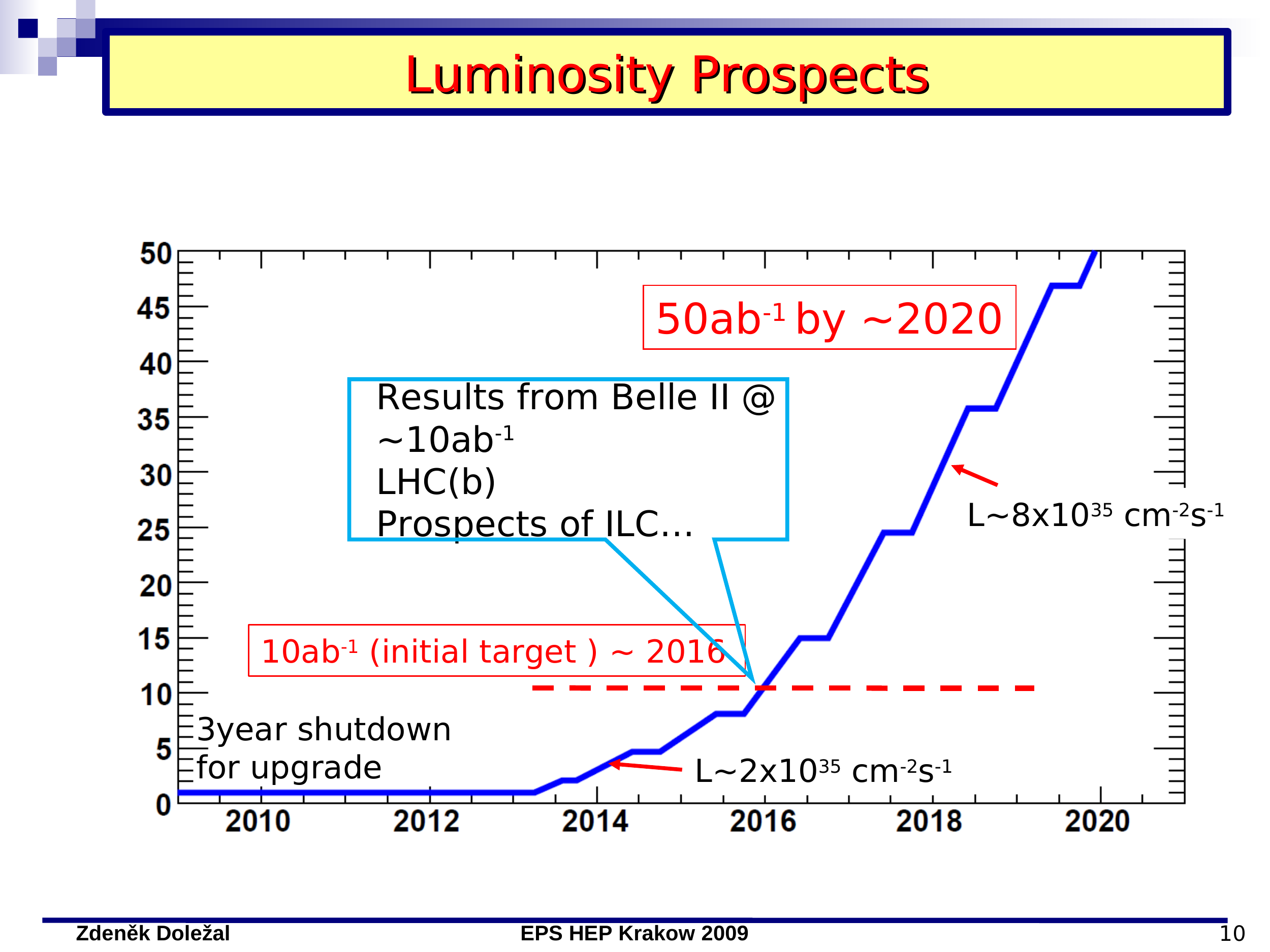}
  \caption{SuperKEKB luminosity prospects}
\label{fig:ramp}
\end{figure}

However the upgrade plans are much more ambitious (see
Fig.~\ref{fig:ramp}). After the 3-year shutdown and obvious learning
curve the upgraded collider
should finally reach  
%The first stage of the upgrade should take place
%during the 3-years shutdown. 
%A factor of 10 increase of instanteneous
%luminosity is planned ($2 \times
%10^{35}\mathrm{cm}^{-2}\mathrm{s}^{-1}$), which should accumulate into
%10 ab$^{-1}$
%by 2016. Then the strategy will be reviewed, taking into account 
%also results from LHC experiments. The second stage should bring the
%luminosity up to     
$8 \times
10^{35}\mathrm{cm}^{-2}\mathrm{s}^{-1}$ -- 40 times higher than today.
This will allow to collect an integrated luminosity of 50 ab$^{-1}$
by 2020.

How can these values be achieved? 
Luminosity can be expressed in a following form:
\begin{equation}
\mathcal{L}=\frac{\gamma_{\pm}}{2er_e}
(1+\frac{\sigma^*_y}{\sigma^*_x})\frac{I_{\pm}\xi_{\pm}R_L}{\beta_y^*
  R_y}
\end{equation}
Here $\gamma_{\pm}$ are relativistic factors of electron and positron
beam, $r_e$ is the classical electron radius, $ \sigma^*_x$ and
$\sigma^*_y$ are beam dimensions at the interaction point (IP) in horizontal and vertical
directions.  $I_{\pm}$ are the currents of both beams, $\xi_{\pm y}$
is the beam-beam  parameter, $\beta_y^*$ is the vertical $\beta$
function at the IP and ($R_L/R_y$) is a luminosity reduction factor
(or tune shift) which reflects crossing at the finite (non-zero)
angle.

To increase luminosity two options have been studied at KEK. One,
called \emph{high-current option} requires increase of stored beam
current from 1.7/1.4~A to 9.4/4.1~A. At the same time  the
beam-beam parameter has to be increased from 0.1 to 0.3 or more.
Recent simulations and tests show, that this option will not deliver
the luminosity expected. Therefore, a work on the second option, called
\emph{nano-beam} has started. This is based on the idea of P.
Raimondi from Frascati~\cite{raimondi} and assumes drastic squeezing the
beam size at the IP.   This means lowering $\beta_y^*$ from current
value of 6~mm down to 0.27/0.42~mm and also slight increase of currents to
3.6/2.6~A. 
The comparison of preliminary parameters for both options as well as
for the current machine is shown in Table~\ref{tab:machinepar}.

 \begin{table}
  \centering
\begin{tabular}{|l|l|l|l|l|}
\hline
\bf Parameter & \multicolumn{2}{c|} {\bf KEKB} & \multicolumn{2}{c|}
    {\bf SuperKEKB}   \\
\hline
LER/HER & design & achieved & high-current & nano-beam \\
\hline
$\beta_y^*$ [mm] & 10/10 & 5.9/5.9 & 3/6 & 0.27/0.42 \\
\hline
$\sigma^*_y$ [$\mu$m]& 1.9 & 1.1 & 0.85/0.73 & 0.084/0.072\\
\hline
$\xi_{\pm y}$ & 0.052 & 0.101/0.096 & 0.3/0.51 & 0.088/0.09 \\
\hline
$I_{\pm}$ [A] & 2.6/1.1 & 1.62/1.15 & 9.4/4.1 & 3.6/2.6\\
\hline
$L$ [ $10^{34}\mathrm{cm}^{-2}\mathrm{s}^{-1}$] & 1 & 2.1 & 53 & 80\\
\hline
\end{tabular}
\caption{Machine parameters for various options (preliminary). For the
notation see text}
\label{tab:machinepar}
\end{table}
\section{Detector Upgrade}
The increase of the luminosity will inevitably bring higher occupancy
and radiation doses. To cope with them an upgrade of Belle detector is
planned. Large number of the detector components will be reused, but many
others have to be replaced to prevent compromising the detector performance.

The upgrade of Belle is pursued by a new collaboration called
Belle~II~\cite{b2www}. This collaboration consisting of around 45
institutes is represented by a spokesperson
P. Kri\v{z}an from Ljubljana and the Institute and Executive Boards. The regular collaboration meetings are
open and new collaborators are welcome.

Vertex detector, built from the double-sided silicon strip sensors
will be replaced by 2 layers of pixels based on DEPFET technology (see
next section). They will be followed by four layers of  double-sided
silicon strip sensors (DSSD). To cope with high occupancy the strips will be
read out by fast APV25 chips, designed for the CMS experiment at the LHC. The
outer radius of silicon will extend to 14~cm.

A new drift chamber with higher granularity and He/C$_2$H$_6$ is under
design to measure tracks and d$E$/d$x$. 

The key requirement for detectors at Super B factories is an ability
to separate kaons from pions. This will be provided by
Time-of-Propagation counter in the barrel region and proximity
focusing Cherenkov ring imaging counter with aerogel radiators in the
endcap. TOP will measure the time that the internally reflected light
travels down the quartz bar and one spatial coordinate along the bar.

The main concern for the electromagnetic calorimeter (ECL) is the
background increase. Hence the readout electronics will be replaced by
a new verson equipped with pipeline and waveform sampling. CsI(Tl) crystals
in the barrel will remain the same, but pure CsI will be used in the
endcap. 

The muon system of Belle based on resistive plate chambers will
probably remain unchanged in the barrel region. The harsh background
conditions in the endcap region will force us to use a scintillator based
solution, with the  light detected by the Geiger mode APDs.

A new readout system is under design. The main chalenge is to handle
large data volumes from the pixel detector.

Individual detector components are prototyped and technology choices
will be made in the first half of 2010. The detailed description of
the detector upgrade can be found in~\cite{detstudy}.
 
\section{DEPFET pixel vertex detector}
The most chalenging experimental requirement is the detection of the
decay point of the short-living B-mesons, relying on  high-performance
vertex detector. Therefore two inner layers of Belle II vertex detector will
use DEPFET pixels. 

The DEPleted Field Effect Transistor structure,  provides detection and amplification properties
jointly.  A MOS
or junction field effect transistor is integrated onto a detector
substrate. By means of sidewards depletion, appropriate bulk,
source and drain potentials, and an additional deep-n-
implantation, a potential minimum for electrons is created right
underneath the transistor channel.
This can be regarded as an internal transistor gate. A
particle entering the detector creates electron-hole pairs in the
fully depleted silicon substrate. While the holes drift to the rear
contact of the detector, the electrons are collected and stored in the
internal gate. The signal charge
leads to a change in the potential of the internal gate, resulting
in a modulation of the channel transistor current. After
readout, the signal charges are cleared out of the internal gate.
A low noise is obtained because of the small capacitance of the
internal gate. Since all charge
generated in the fully depleted bulk is collected, the device
provides a high signal as well. Both together yield a very large
S/N ratio. The DEPFET pixel detectors have been developed for vertex
detection at the ILC for many years and  have now received high
level of maturity~\cite{depfet2}. Several beam tests have shown that
pixels of $24\times38 \mu$m$^2$ pitch achieve intrinsic
resolution below 2~$\mu$m. 

For the Belle II vertex detector the use of very thin
(50 $\mu$m) detectors is planned with the pitch around
$38\times50 \mu$m$^2$. The detector geometry is still
under discussion, but preliminary radii of the two layers are 1.3 and
2.2~cm. Simulations show, that use of pixels brings a factor of 3
improvement in an impact parameter resolution when compared to
originally proposed DSSD.
It is expected that the vertex detector will be delivered by the
DEPFET collaboration centered in Europe (Czech Republic, Germany,
Poland, Spain). 

\section{Conclusions}
As shown in the previous sections, exciting physics experiment is
under preparation at KEK. Both accelerator and detector upgrade proposals are
well under way. 
%The funding applications have been submitted and the
%outcome should be known by the end of 2009. 
An open international
collaboration has been created.

\end{document}